\documentclass[a4paper,usenatbib,onecolumn]{mn2e}
\usepackage{newtxtext,newtxmath}
\usepackage{lipsum}
\usepackage[T1]{fontenc}
\usepackage{ae,aecompl}
\usepackage{hyperref}
\usepackage{graphicx}	% Including figure files
\usepackage{amsmath}	% Advanced maths commands
\usepackage{amssymb}	% Extra maths symbols
\usepackage{bm}	% Extra maths symbols
\usepackage{caption}
\usepackage{multirow}
\usepackage{multicol}
\usepackage{cuted}
\usepackage{mathtools}
\usepackage{longtable}
\usepackage{rotating}

\title[Fomalhaut]{Understanding Fomalhaut as a Cooper pair}
\author[F. Feng et al.]
{F. Feng$^{1}$\thanks{E-mail: f.feng@herts.ac.uk or fengfabo@gmail.com}, H. R. A.  Jones$^{1}$\\
$^{1}$Centre for Astrophysics Research, University of Hertfordshire, College 
Lane, AL10 9AB, Hatfield, UK
}
\date{\today}

\begin{document}
\maketitle

\begin{abstract}
Fomalhaut is a nearby stellar system and has been found to be a triple based on astrometric observations. With new radial velocity and astrometric data, we study the association between Fomalhaut A, B, and C in a Bayesian framework finding that the system is gravitationally bound or at least associated.  Based on simulations of the system, we find that Fomalhaut C can be easily destabilized through combined perturbations from the Galactic tide and stellar encounters. Considering that observing the disruption of a triple is probably rare in the solar neighbourhood, we conclude that Fomalhaut C is a so-called ``gravitational pair'' of Fomalhaut A and B. Like the Cooper pair mechanism in superconductors, this phenomena only appears once the orbital energy of a component becomes comparable with the energy fluctuations caused by the environment. Based on our simulations we find (1) an upper limit of 8\,km/s velocity difference is appropriate when selecting binary candidates and (2) an empirical formula for the escape radius, which is more appropriate than tidal radius when measuring the stability of wide binaries.
\end{abstract}
%%%%%%%%%%%%%%%%%%%%%%%%%%%%%%%%%%%%%%%%%%%%%%%%%%%%%%%%%%%%%%%%%%%%%%%%%%%%
\section{Introduction}     \label{sec:introduction}
Fomalhaut ($\alpha$ PsA, HD216956, HIP 113368) is a nearby A3V-type star, hosting a debris disk and a candidate planet which may have shaped the structure of the disk \citep{kalas05,kalas13}. Despite optical detection the candidate planet has eluded detection in the infrared \citep{janson12} and so a number of hypotheses have been investigated - even that it is a plausibly cool neutron star \citep{poppenhaeger17}. According to \cite{matra17}, CO emission has been detected from the Fomalhaut disk, indicating a release of exocometary gas or the occurrence of a recent impact event. 

TW PsA (HIP 113283, hereafter Fomalhaut B) and LP 876-10 (hereafter Fomalhaut C) were claimed to be bound to Fomalhaut (hereafter Fomalhaut A) because they share common parallax, proper motion and radial velocity. Fomalhaut B's companionship and coevality was established by \cite{navascues97} and Fomalhaut C's companionship based on \cite{mamajek12} and \cite{mamajek13}, hereafter M12 and M13 respectively. The masses of Fomalhaut A, B, and C are 1.92$\pm$0.02, 0.73$^{+0.02}_{-0.01}$, and 0.18$\pm$0.02\,$M_\odot$, respectively. Fomalhaut B and C are currently located about 0.28 and 0.77\,pc away from Fomalhaut A, respectively (M12,M13). 

M12 find a 0.1$\pm$0.5\,km/s difference between the peculiar velocities of A and B while M13 find a $\sim$1\,km/s difference between AB and C. The association of C to AB is also supported by the distances from C to A and B being about 0.77$\pm$0.01\,pc and 0.987$\pm$0.006\,pc, respectively. These separations are allowed by the tidal radius of the system, which is 1.9\,pc according to \cite{jiang10}. Aside from the astrometric arguments about the coeval nature of the components M13 consider age and metallicity. In particular, the age of C is not well determined due to uncertainties in the pre-main-sequence evolution tracks in the colour-magnitude diagram (see fig. 4 of M13). 

Fomalhaut C shares a common velocity with AB within 1.1$\pm$0.7\,km/s, slightly larger than the escape velocity of 0.2\,km/s (M13). M13 claim that it is rare to find an M-dwarf in the vicinity of AB with such a low velocity difference. They also consider that C should not be escaping AB since it is unlikely to catch a temporary phenomena in the act. To investigate the multiplicity of Fomalhaut, we use the archived astrometric data and HARPS radial velocities to determine the kinematics of Fomalhaut. We study the probability of the associations between A, B, and C using analytical, statistical, and numerical methods. Assuming the multiplicity of Fomalhaut, we investigate the stability, formation and evolution of this system based on simulations of clones of the system under the perturbations from the Galactic tide and stellar encounters. We also discuss the hypothesis that the eccentricity of the disk around Fomalhaut A is caused by dynamical perturbations from Fomalhaut B and C \citep{shannon14,kaib17}. 

The paper is structured as follows. First, we describe the data in section \ref{sec:data}. We then study the problem analytically and statistically using both the new and old astrometric data in section \ref{sec:analytical}. We simulate the Fomalhaut system in a Monte Carlo fashion in section \ref{sec:perturb} and find an empirical formulae for the escape radius in section \ref{sec:radius}. Finally, we draw conclusions in section \ref{sec:conclusion}. 

\section{Data}\label{sec:data}
% In the catalogue of close encounter pairs (CEP1, \citealt{feng17c}), we find three encounters of Fomalhaut A and B based on reliable data. They are HIP 85397, HIP 61743, LAMOST J081618.61+053754.6, passing Fomalhaut A at\footnote{The optimal value is the nominal one while the lower and upper errors are defined by the 5\% and 95\% quantiles of the distribution of 1000 clones of encounters. } $0.57_{-0.05}^{+0.41}$, $0.19_{-0.00}^{+2.33}$, and $0.84_{-0.15}^{+3.39}$\,pc at 1.59, 2.82, and -0.32\,Myr, respectively. HIP 61743 and HIP 85397 have masses of 0.87 and 1.40 \,$M_\odot$ \citep{ramirez12,chandler15}, respectively.

In Table \ref{tab:data} we show the data that we have used for this study.
\begin{table*}
  \centering
  \caption{Astrometric data of Fomalhaut A, B and C in M13 and in this work (F17). The references (Ref.) for the data used in F17 are denoted by 1, 2, 3, 4, 5, and 6, corresponding to \protect\cite{leeuwen07}, \protect\cite{weinberger16}, \protect\cite{gontcharov06}, M13, F17, and \protect\cite{zacharias17}, respectively. $\Delta_{\rm com}$ is the distance from a component to the system's center of mass. $\Delta S$ is the velocity of Fomalhaut B or C to Fomalhaut A.} 
  \label{tab:data}
  \begin{tabular}{ccccccccc}
    \hline 
    Parameter & \multicolumn{2}{c}{A} & \multicolumn{2}{c}{B}  & \multicolumn{2}{c}{C}  & Unit & Ref.\\
    &M13&F17&M13&F17&M13&F17&&\\
    \hline
    $\alpha_{\rm ICRS}$(J2000)&344.411773&344.412693&344.099277&344.100222&342.018632&342.0186992&deg&1,1,6\\
    $\delta_{\rm ICRS}$(J2000)&-29.621837&-29.621608&-31.565179&-31.564958&-24.368872&-24.3687906&deg&1,1,6\\
    $\pi$&129.81$\pm$0.47&129.81$\pm$0.47&131.42$\pm$0.62&131.42$\pm$0.62&132.07$\pm$1.19&129.57$\pm$0.31&mas&1,1,2\\
    $\mu_\alpha$&328.95$\pm$0.50&328.95$\pm$0.50&331.11$\pm$0.65&331.11$\pm$0.65&333.8$\pm$0.5&327.1$\pm$0.90&mas\,yr$^{-1}$&1,1,6\\
    $\mu_\delta$&-164.67$\pm$0.35&-164.67$\pm$0.35&-158.98$\pm$0.48&-158.98$\pm$0.48&-177.5$\pm$0.7&-183.5$\pm$0.90&mas\,yr$^{-1}$&1,1,6\\
    $v_r$&6.5$\pm$0.5&6.5$\pm$0.5&6.6$\pm$0.1&7.222$\pm$0.013&6.5$\pm$0.5&6.5$\pm$0.5&km\,s$^{-1}$&3,5,4\\
    $X_{\rm gal}$&3.06&3.06&3.14&3.14&3.01&3.07&pc&5,5,5\\
    $Y_{\rm gal}$&1.14&1.14&0.90&0.89&1.86&1.89&pc&5,5,5\\
    $Z_{\rm gal}$&-6.98&-6.98&-6.88&-6.88&-6.70&-6.82&pc&5,5,5\\
    $U$&-5.71$\pm$0.16&-5.72$\pm$0.21&-5.69$\pm$0.06&-5.44$\pm$0.08&-5.34$\pm$0.19&-5.67$\pm$0.19&km\,s$^{-1}$&5,5,5\\
    $V$&-8.26$\pm$0.28&-8.26$\pm$0.08&-8.16$\pm$0.07&-8.08$\pm$0.05&-7.58$\pm$0.28&-8.22$\pm$0.08&km\,s$^{-1}$&5,5,5\\
    $W$&-11.04$\pm$0.38&-11.04$\pm$0.47&-10.96$\pm$0.08&-11.52$\pm$0.14&-11.12$\pm$0.42&-11.83$\pm$0.45&km\,s$^{-1}$&5,5,5\\ 
    $\Delta_{\rm com}$&0.05&0.04&0.24&0.25&0.77&0.77&pc&5,5,5\\
    $\Delta S$&0&0&0.13$\pm$0.51&0.59$\pm$0.43&1.12$\pm$0.72&0.11$\pm$0.06&km\,s$^{-1}$&5,5,5\\\hline
  \end{tabular}
\end{table*}
We use astrometric parameters from \cite{leeuwen07} and astrometric measurements of Fomalhaut C from \cite{zacharias17} and \cite{weinberger16}. These values do not indicate a simple overlap of distance and proper motion between the A, B and C components. M12 and M13 have used revised Hipparcos (dubbed HIP2) position without converting to epoch J2000. We correct this and show the HIP2 position at J2000 in the table. We have also corrected a typo in M13 which gives $\mu_\alpha=329.95$ rather than 328.95\,mas/yr in the HIP2 catalog. 

M12 and M13 use the radial velocity value of 6.5$\pm$0.5 for A from \cite{gontcharov06} determined from older literature sources and 6.6$\pm$0.1 for B based on \cite{nordstrom04} who use seven points from Coravel taken over 3794 days. Other literature values from modern radial velocity spectrographs have become available. For example, those of 7.76 and 7.78\,km/s for A from CORALIE by \cite{erspanmer03} and 7.217$\pm$0.0163 for B from CORALIE and HARPS from \cite{soubiran13} based on four CORALIE and six HARPS points taken over 2500 days. Using the HIRES spectrograph \citep{chubak11} report a mean radial velocity for B of 7.152$\pm$0.086 based on four measurements over 1619 days. \cite{lagrange13} also report on taking 284 radial velocities of A over 1853 days with HARPS to look for radial velocity variations. These include high frequency radial velocity measurements which indicate an amplitude variation of some 300\,m/s over the course of an hour an half. The measurements show a radial velocity amplitude of 620\,m/s over the course of their observations with no reported periodic radial velocity signal or trend in their relative radial velocities. For the C component there appears to be only one literature measurement for the radial velocity of C, with M13 reporting a value of 6.5$\pm$0.5\,km/s based on a comparison between CRIRES spectra and a synthetic spectrum. 

We also find values from the HARPS archive for A, B and C. There are 1100 radial velocity points for A taken over 11 years, 216 for B taken over year and two exposures for C taken within an hour on 2014 November 21. However the pipeline absolute radial velocities for A and C are not particularly useful since the default G2 template is poorly matched to an A star and the derived values are not stable. For C the radial velocities are problematic since they have a HARPS archive signal-to-noise of 3.7 and 4.6 and show very significant scatter with values of 5.783$\pm$0.040 and 6.903$\pm$0.026\,km/s.  Given the variability and uncertainties on the radial velocities and that our results are not sensitive to the different values we choose to use the radial velocities from M13 for A and C, and the radial velocity of B based on the HARPS radial velocities. The values adopted are in Table \ref{tab:data}. Since the radial velocity relative errors of the Fomalhaut A, B, and C are around 1\,km/s, it is not necessary to account for the convective blueshift and gravitational redshift, which typically cause a relative radial velocity difference less than 1.0\,km/s \citep{reiners15,pasquini11}. 

\section{Bayesian Inference for the association of Fomalhaut components}\label{sec:analytical}
Instead of identifying wide binary candidates by adopting an arbitrary threshold, we apply Bayesian statistics to quantify the probability of A, B, and C to be a bound system. We test the hypotheses of binary ($H_1$) and non-binary ($H_0$) for a given data set. According to the Bayes theorem, the posterior ratio of the two hypotheses is
\begin{equation}
  \frac{P(H_1|D)}{P(H_0|D)}=\frac{P(D|H_1)}{P(D|H_0)}\frac{P(H_1)}{P(H_0)}~,
\end{equation}
where $D$ is the data, $P(D|H_1)$ and $P(D|H_0)$ are the evidences, and $P(H_1)$ and $P(H_0)$ are the priors of $H_1$ and $H_0$, respectively. According to \cite{raghavan10}, 56\% solar-type stars are single. Thus it is reasonable to assume that the single and binary hypotheses are equally probable, $i.e. P(H_1)=P(H_0)=0.5$. Then the posterior ratio becomes the evidence ratio or Bayes factor which is
\begin{equation}
 {\rm BF}_{10}=\frac{P(D|H_1)}{P(D|H_0)}~.
\end{equation}
If ${\rm BF}_{10}>150$, we claim that $H_1$ is favoured by the data (e.g., \citealt{kass95}).

Considering that $P(D)=P(H_1|D)+P(H_0|D)=1$, the evidence of $H_1$ is
\begin{align}
  P(D|H_1)=\frac{P(H_1|D)P(D)}{P(H_1)}=2P(E<0|E_{\rm obs},\sigma_E)~, 
\end{align} 
where $E$ is the orbital energy of the target star pair, and $E_{\rm obs}$ and $\sigma_E$ are the observed orbital energy and corresponding error for the test star. We calculate $P(E<0|E_{\rm obs},\sigma_E)$ by drawing a sample of clones of the test star according to the data uncertainty and calculating the proportion of the bound clones in the whole sample. Then we derive the evidence for $H_0$, i.e. $P(D|H_0)$. We assume the encounter velocity follows the Maxwell-Boltzmann distribution with a mean of 75\,km/s (based on \citealt{feng17c}), the probability of measuring an encounter velocity less than a threshold $v_{\rm th}$ is
\begin{equation}
  P(v<v_{\rm th}|H_0)= {\rm erf}\left( \frac{v_{\rm th}}{\sqrt{2}a}\right)-\sqrt{\frac{2}{\pi}}\frac{v_{\rm th}e^{-v_{\rm th}^2/(2a^2)}}{a}~,
  \label{eq:pv}
\end{equation}
where ${\rm erf}$ is the error function, and $a=47$\,km/s is the scale velocity. For example, $P(v<1~{\rm km~s}^{-1}|H_0)=2.56\times10^{-6}$. For a sample of $N$ stars within a distance of $d_{\rm max}$ from the Sun or from an arbitrary reference point, the probability of finding two stars separated by less than $d_{\rm th}$\footnote{The equation only applies for $d_{\rm th}$ far less than the average separation between stars. For large $d_{\rm th}$, the probability becomes number density and could be larger than one. } is
\begin{equation}
  P(d<d_{\rm th}|H_0)= N\left(\frac{d_{\rm th}}{d_{\rm max}}\right)^3~.
    \label{eq:pd}
\end{equation}
For example, the probability of finding two stars separated by 1\,pc in the 8\,pc-sample of 211 stars in the solar neighborhood \citep{kirkpatrick12} is about 0.4. 

Combining Eqs. \ref{eq:pv} and \ref{eq:pd}, we derive the evidence of $H_0$, which is 
\begin{equation}
P(D|H_0)=P(v<v_{\rm th}|H_0)P(d<d_{\rm th}|H_0)~.
\end{equation}
If Fomalhaut A and B are not bound or associated, the probability of observing their current position and velocity difference (see the F17 data in Table \ref{tab:data}) in the 8\,pc sample \citep{kirkpatrick12} is $5.3\times10^{-9}$. If A, B, and C are randomly distributed in the phase space, the probability of observing their current configuration is $3.4\times10^{-18}$. Comparing the contribution of spatial and velocity differences to the evidence, we find that the velocity difference is more crucial in the detection of wide multiples. For example, the probability of finding a velocity difference of 4\,km/s between two field stars is $1.64\times10^{-4}$. Considering an event with a probability of $10^{-3}$ as rare, we suggest an upper limit of velocity difference of 8\,km/s for stars separated by less than 1\,pc as a reliable criterion in the search for wide binaries. 
  
\begin{figure*}
  \centering 
  \includegraphics[scale=0.7]{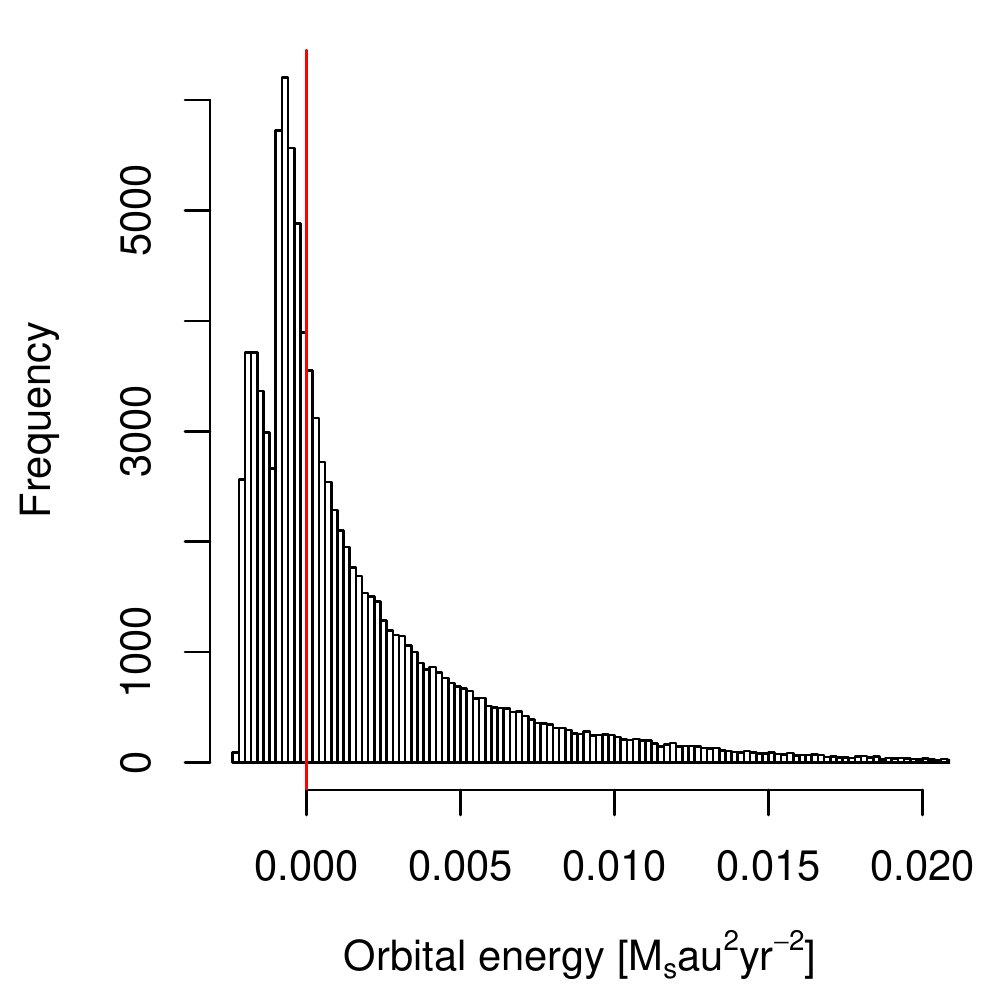}
  \includegraphics[scale=0.7]{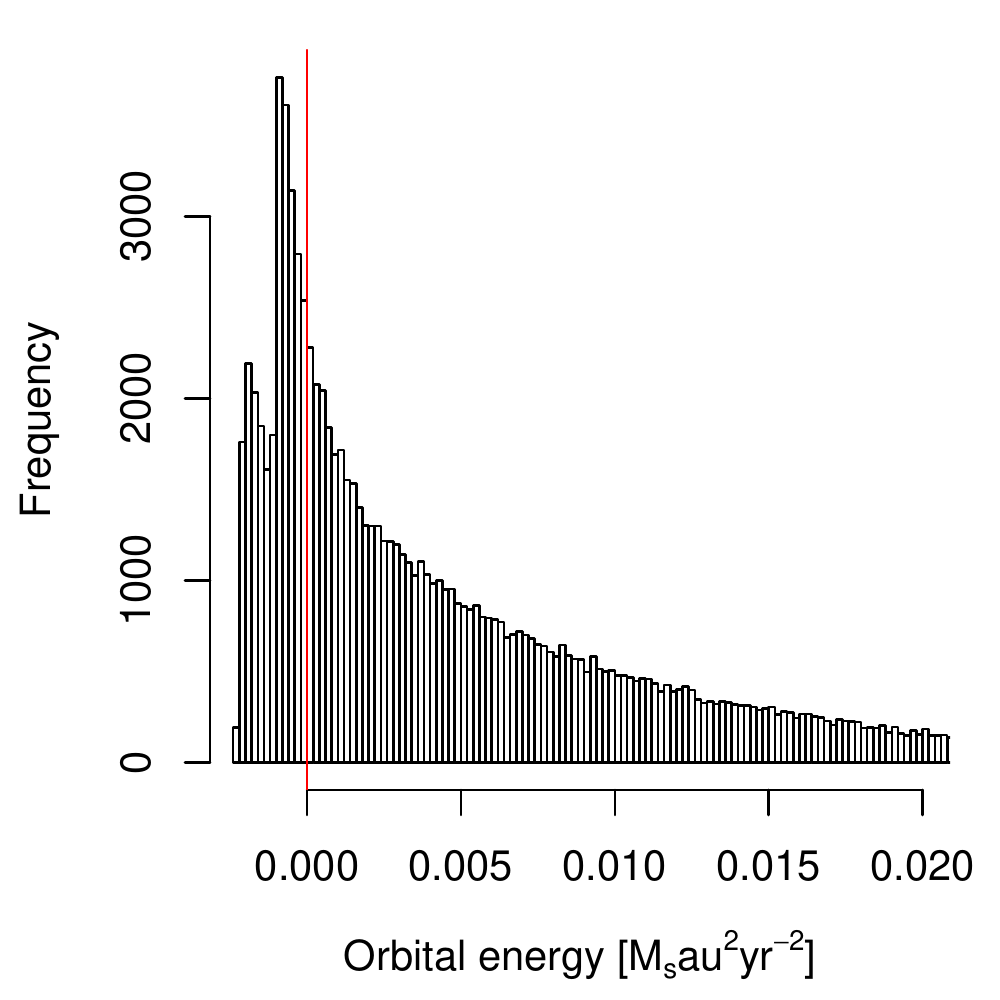}
  \caption{Distribution of orbital energy of $10^5$ clones of Fomalhaut A and B based the M13 (left) and F17 (right) data. The red line denotes the zero orbital energy. The clones with orbital energy less than zero are counted as bound clones. Otherwise, they are unbound clones.}
  \label{fig:dv}
\end{figure*}

To calculate the evidence for $H_1$, $P(D|H_1)$, we generate clones of B according to the M13 data listed in Table \ref{tab:data}. We draw one million clones from Gaussian distributions with means and standard deviations from the astrometric parameters in table 1 of M13. We show the distribution of orbital energy of B and A in Fig. \ref{fig:dv}. We find a probability of 0.54 for A and B to form a binary based on $10^5$ clones in addition to the nominal astrometric values of A and B. Assuming nominal parameters for A and B, we further investigate the association of C to AB with a similar method and find a probability of 0.45 that the orbits are bound. 

We also investigate the association of A, B, and C using the F17 data in Table \ref{tab:data}. Then we find a probability of 0.31 and 0.27 for B and C to be bound or associated to A, respectively. Therefore Fomalhaut as a triple system is favoured by the data since ${\rm BF}_{\rm 10}=P(D|H_1)/P(D|H_0)= 7.9\times10^{16}\gg 150$. 

\section{Stability of Fomalhaut under perturbations from the Galactic tide and encounters}\label{sec:perturb}
It is also plausible that the Fomalhaut triple formed together but was disrupted to be a stellar association, as indicated by the recent discovery of wide binaries from Gaia DR1 \citep{price-whelan17}. We generate clones of the Fomalhaut system and study the dynamical stability of these clones based on simulations. We use the numerical methods of \cite{feng14} and \cite{feng18} to simulate their motions under the perturbations from the Galactic tide and stellar encounters with a 1\,kyr time step. Following \cite{feng18}, we simulate 80 encounters with periapsis less than 1\,pc every Myr. We then simulate the orbits of the clones with perturbations from encounters which have encounter time close to each time step. Thus multiple encounters can be accounted for simultaneously. Considering that the encounter time scale is much shorter than the orbital period of clones, we use the impulse approximation to calculate the velocity kicks from encounters \citep{feng18}. 

To compare the role of the Galactic tide and stellar encounters in perturbing binaries, we simulate the motions of A and B under the perturbations from the Galactic tide alone (AB+T simulations) and from both the tide and encounters (AB+TE simulations). We also perform these two types of simulations for A, B, and C (i.e., ABC+T and ABC+TE simulations). We derive initial conditions of A, B, and C from the parameters listed in Table \ref{tab:data} and select 200 clones (sometimes 500 clones) which are stable over 1\,Myr simulation under the perturbations from the Galactic tide. For each set of simulations, we integrate the orbits of two hundred clones backwards from the present to 500\,Myr ago.

Following \cite{feng18}, we count a clone of B as unstable if its eccentricity is larger than 1 by the end of simulations. Considering the non-Keplerian motion of C, we count a clone of C as unstable if its orbital energy with respect to A and B is larger than zero. We call this the ``relaxed criterion'' since the stability is determined only by the final state. We also define the ``exclusion criterion'' by counting the clones which have eccentricity larger than 1 or orbital energy larger than zero at any time during the simulations. As concluded by \cite{kaib17}, the motion of C is not Keplerian while the orbit of B is approximately Keplerian because C is less massive and further than B with respect to A. The eccentricity of B is calculated with respect to A while the orbital energy of C is calculated with respect to the barycenter of A and B. Considering that many clones on highly eccentric orbits may be brought back to low eccentricity orbits through perturbations of the Galactic tide and encounters, we use our ``relaxed'' criterion to determine the instability of a clone by default. 

\begin{figure*}
  \centering 
  \includegraphics[scale=0.5]{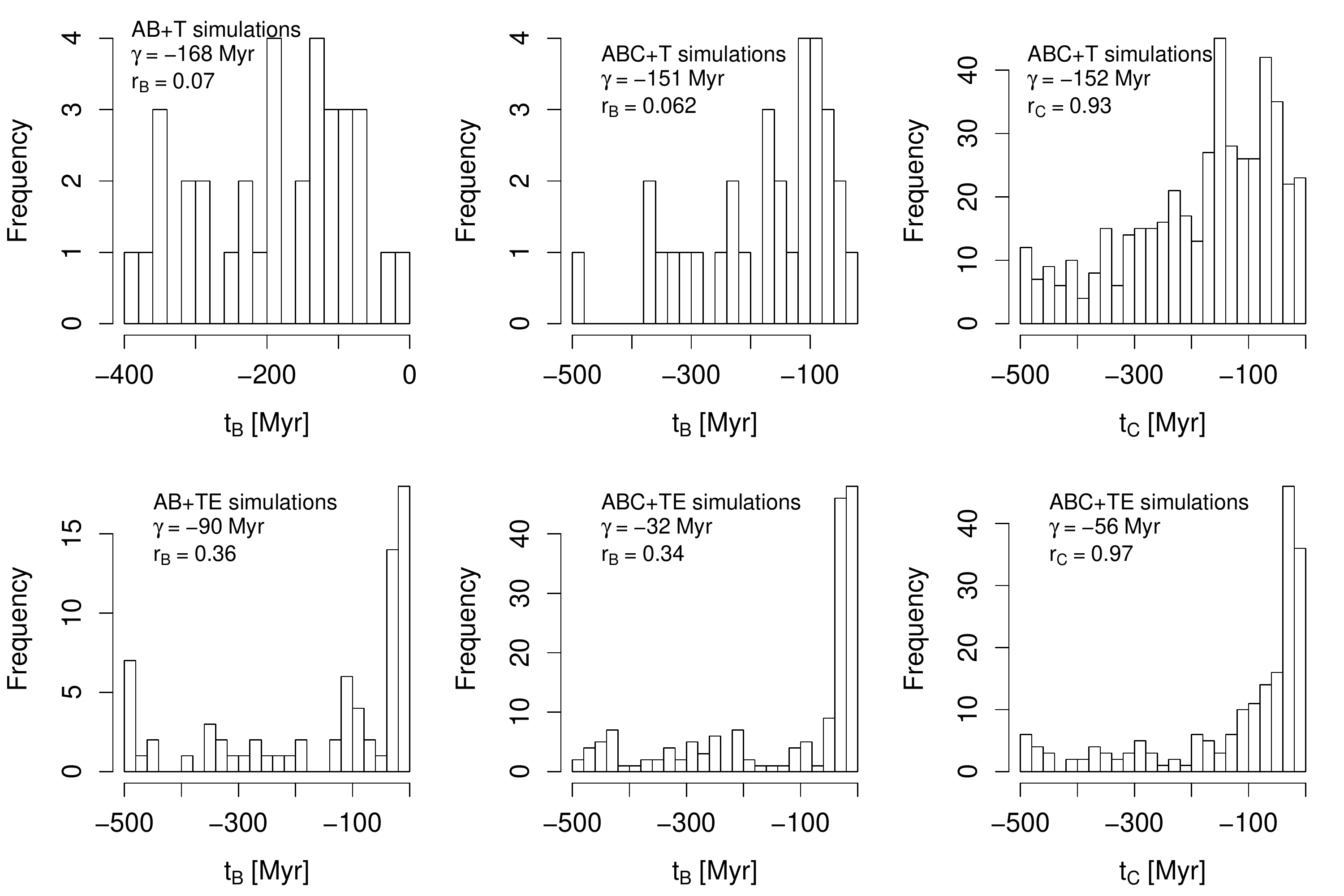}
  \caption{Distribution of the disruption time of AB (left and middle panels), and AC (right panels) in various sets of simulations. The median of the disruption time ($\gamma$) is shown in the top left corner of each panel. The disruption/ejection ratio of AB ($r_B$) and AC ($r_C$) are also shown.}
  \label{fig:disruption}
\end{figure*}

The ejection ratio and the distribution time of the ejected clones for different sets of simulations are shown in Fig. \ref{fig:disruption}. We find that the ejection ratio of B is always lower than C for all simulations because AB is more tightly bound than AC. The ejection ratio of B is higher in the TE simulations than in the T simulations because the combined perturbation of the tide and encounters are stronger than that of the tide alone. We also observe that the ejection ratio of C in ABC+TE simulations is slightly higher than in ABC+T simulations. For all simulations, we find that encounters significantly reduce the disruption time scale of B and C. This is consistent with the conclusion of \cite{shannon14} that the current orbital configuration of Fomalhaut is probably in a transient stage based on our tide-only simulations. However, the time scale of such a transient phenomenon is likely much less than the age of the system, making the transient hypothesis unlikely (M13). Considering this, \cite{kaib17} conclude that Fomalhaut is currently in a meta-stable stage which can last for a longer time than a transient stage does. 

To investigate whether Fomalhaut is in a stable or unstable stage, we compare the disruption time and ejection ratio based on the ``relaxed'' and ``exclusion'' criterion in Table \ref{tab:criterion}. From the table, we see that different criteria give similar results for AB simulations. We also find that the ejection ratio of B is higher in ABC simulations than in AB simulations, indicating that C tends to destabilize A and B. In particular, all C clones are ejected in all ABC simulations according to our ``exclusion'' criterion while a few percent of C clones are stable based on the ``relaxed'' criterion. That means some unbound clones of C are brought back onto bound orbits and are counted as stable according to our ``relaxed'' criterion. These clones seem to be on meta-stable orbits, as concluded by \cite{kaib17}. 
\begin{table*}
\centering
  \caption{Ejection ratio $r$ and disruption time $\gamma$ for various simulations based on the ``relaxed'' and ``exclusion'' criteria.}
\label{tab:criterion}
\begin{tabular}{lcccc|cc}
&\multicolumn{4}{c}{Stability of B}&\multicolumn{2}{c}{Stability of C}\\
  Simulation type&AB+T&AB+TE&ABC+T&ABC+TE&ABC+T&ABC+TE\\
 \hline 
$r$ (``relaxed'' criterion) &0.07&0.36&0.06&0.34&0.93&0.97\\
$\gamma$ [Myr] (``relaxed'' criterion) &-168&-90&-151&-32&-152&-56\\
$r$ (``exclusion'' criterion) &0.07&0.36&0.24&0.41&1&1\\
$\gamma$ [Myr] (``exclusion'' criterion) &-126&-90&-99&-26&-82&-36\\
 \hline 
\end{tabular}
\end{table*}

\begin{figure*}
  \centering 
  \includegraphics[scale=0.6]{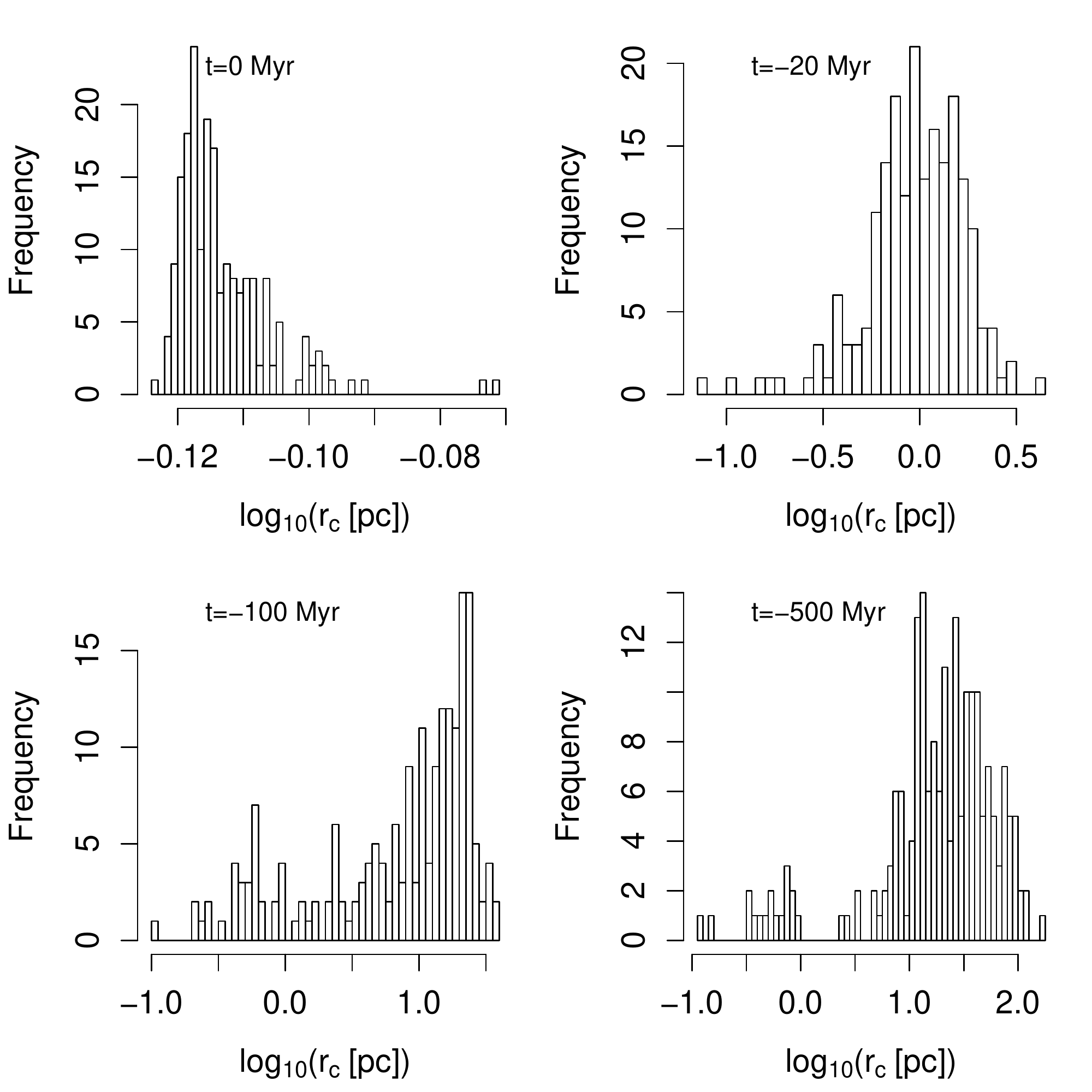}
  \caption{Distribution of the separation between C and A in terms of $r_c$ at 0, -20, -100, and -500\,Myr.  }
  \label{fig:rc}
\end{figure*}

We also test this hypothesis by showing the distribution of the separation between Fomalhaut C and A at different times in the ABC+TE simulations in Fig. \ref{fig:rc}. The initial separations are concentrated around 0.7\,pc. Then most clones of Fomalhaut C become unbound and are around 1\,pc away from A. At $-100$\,Myr, many clones of C move further away from A while some clones remain within 1\,pc of A. Finally, there are two populations of C clones by the end of simulations. Most C clones belong to the first population which is about 10-100\,pc from A. The separation between this population and A will increase with time, as concluded by \cite{jiang10}. On the other hand, there are a few percent of C clones remaining within 1\,pc of A. These clones in the second population do become unbound in the simulations but they are able to remotely associate with A and B because the Galactic tide and encounters may bring them back to the vicinity of A and B.

\begin{figure*}
  \centering 
  \includegraphics[scale=0.6]{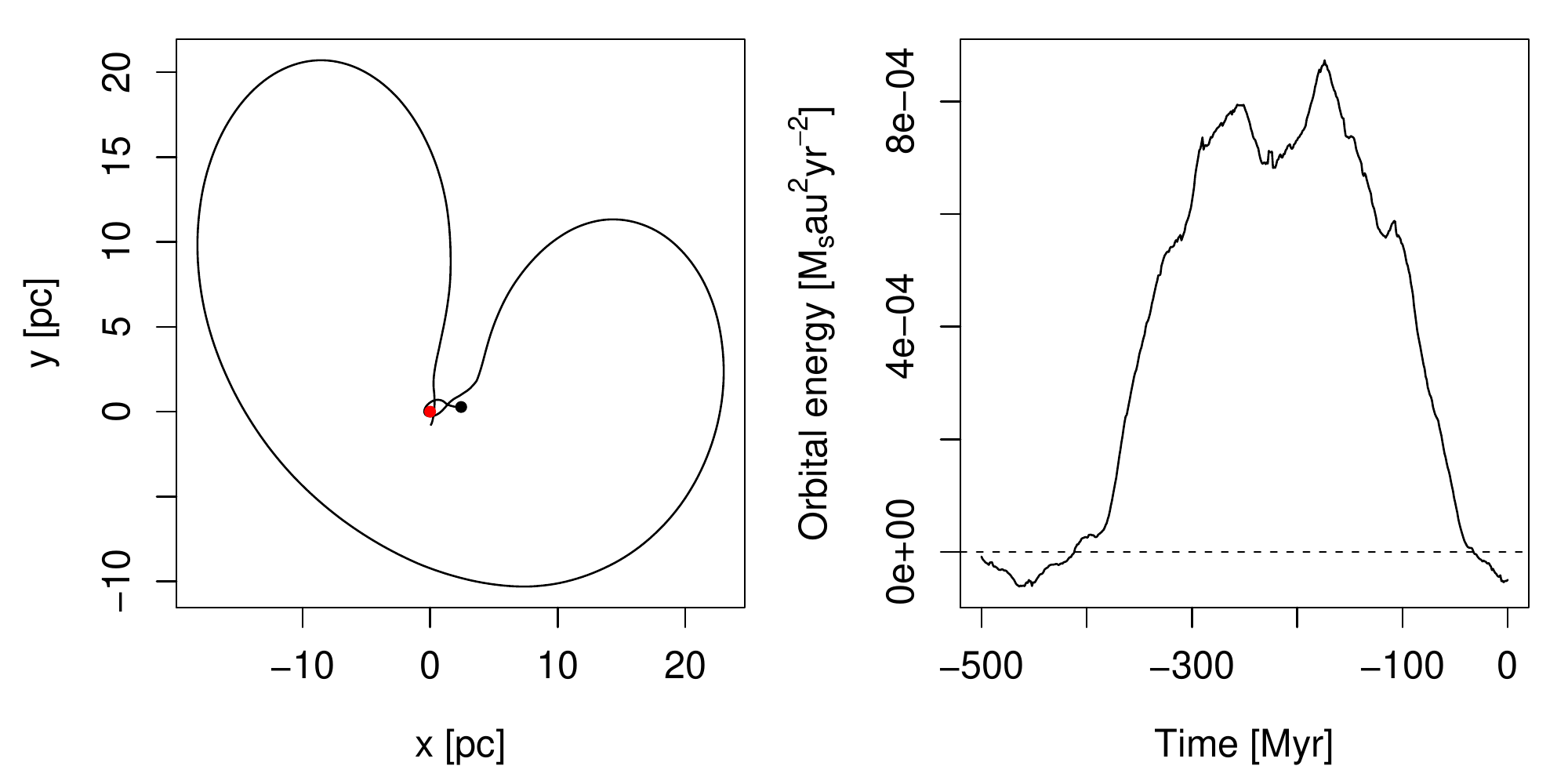}
  \caption{Left panel: orbit of a Fomalhaut C clone with respect to the barycenter (red dot) of Fomalhaut A and B projected onto the Galactic plane. The position at $t=0$\,Myr is denoted by the solid black dot. Right panel: orbital energy of Fomalhaut C with respect to Fomalhaut A and B.}
  \label{fig:pair}
\end{figure*}

We call this population ``gravitational pairs'' of Fomalhaut A and B. These pairs are like the ``Cooper pairs'' in superconductors (Cooper 1956). The pairing between two electrons in a metal becomes possible only if the thermal energy is negligible or the temperature is close to 0 K. Similarly, the gravitational pair only appears when the orbital energy of a wide binary is comparable with the energy fluctuation caused by stellar encounters and the Galactic tide. Thus we conclude that Fomalhaut C is a gravitational pair of Fomalhaut A and B. We illustrate this in Fig. \ref{fig:pair} by showing the orbit of a C clone and the variation of its energy with respect to Fomalhaut A and B. We see that the C clone moves beyond 10\,pc away from A and B and then becomes bound to A and B after about 400\,Myr drift. Since this C clone was bound to the system for about 100\,Myr, Fomalhaut C is more likely to be a gravitational pair of Fomalhaut A and B on a meta-stable orbit rather than to be on a stable (too rare) or unstable orbit (too short-lived). 

\cite{shannon14} and \cite{kaib17} argue that the asymmetry of the disc around Fomalhaut A was caused by the perturbations from the other two components. The orbit of disc objects can be either excited by the instability of the Fomalhaut system \citep{shannon14} or by the periodic perturbations from Fomalhaut B at its periastron which could be as small as 400\,au \citep{kaib17}. We find that about 20\% clones of B come closer than 400\,au to A on the basis that the perihelion of B with respect to A approximately represents the closest distance between them despite perturbations from C. Our results are consistent with the conclusion of \cite{kaib17} that the disc of Fomalhaut A could have been strongly perturbed by the other components and thus became eccentric. 

\section{Empirical formulae of escape radius}\label{sec:radius}
\begin{figure*}
  \centering 
  \includegraphics[scale=0.5]{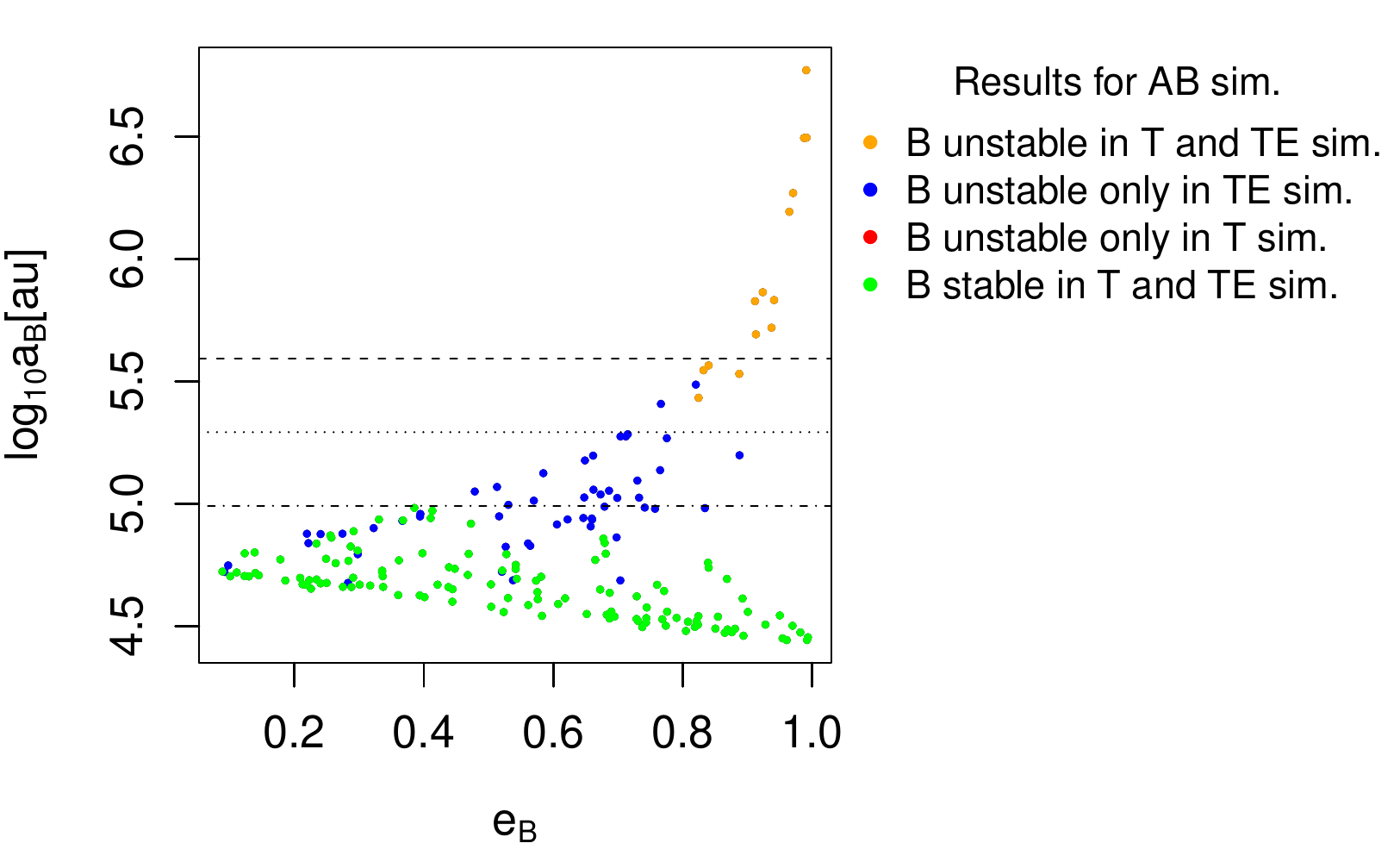}
  \includegraphics[scale=0.5]{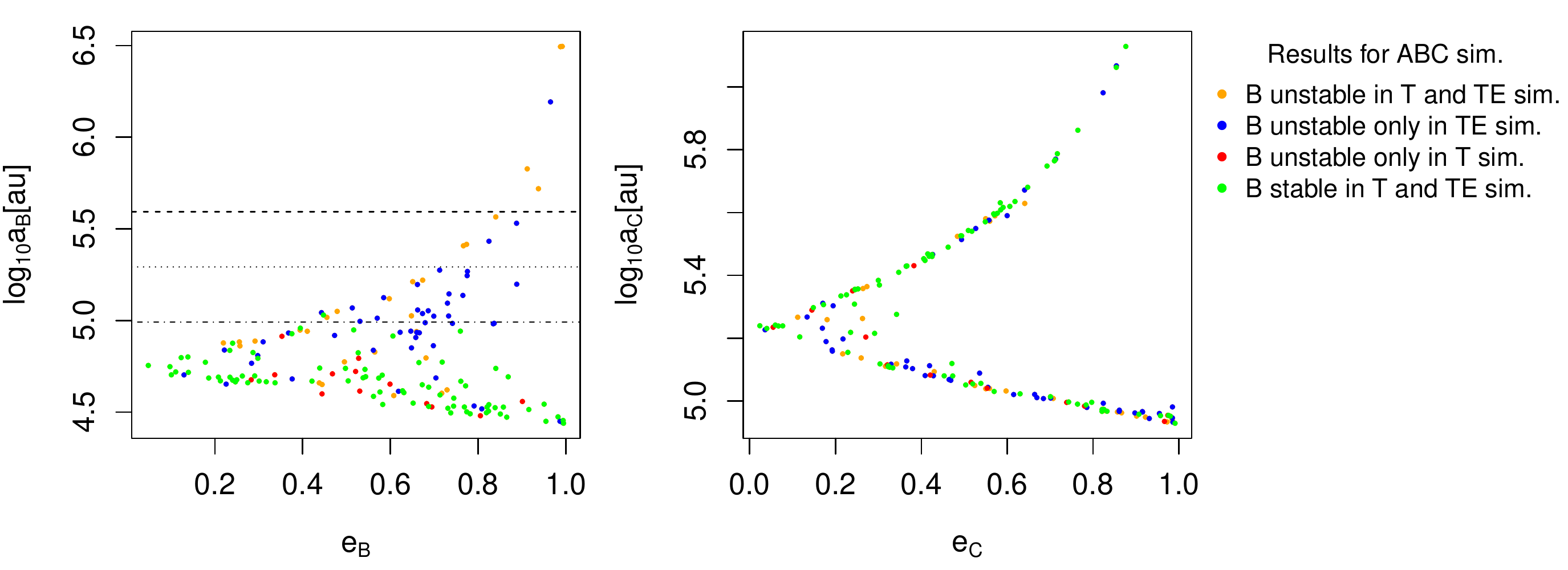}
  \caption{Distribution of initial eccentricity and semi-major axis of clones of AB (upper panel) and ABC (lower panel). In the lower panel the left-hand plot is for $e_B$ versus $a_B$ and the right-hand plot for $e_C$ versus $a_C$ which are calculated with respect to the barycenter of A and B. The horizontal dashed, dotted, and dash-dotted lines represent the tidal radius of AB (1.9 pc) as well as 50\% and 25\% of the tidal radius respectively. These are not shown in the bottom right panel to avoid confusion because the colours encode the stability of A and B while the orbital elements are for C. }
  \label{fig:orbit}
\end{figure*}

To see the correlation between initial orbital elements of clones and their final stages, we show the distribution of initial eccentricity and semi-major axis and encode the final stages of simulations using different colours for different outcomes in Fig. \ref{fig:orbit}. We find that most clones with $a>1$\,pc (about 50\% of the tidal radius of 1.9\,pc calculated by M13 according to \cite{jiang10}) are unstable in the AB+T simulations while most clones with $a>0.5$\,pc (about 25\% of the tidal radius) are unstable in the AB+TE simulations. Orbits with low eccentricity are typically more stable than those with high eccentricity. Therefore the tidal radius is probably not a reliable criterion for stability analysis of wide binaries because it only accounts for perturbations from the Galactic tide. Since \cite{jiang10} do not account for anisotropic encounters and underestimate the encounter rate \citep{feng18}, the tidal radius they have proposed may not be reliable, as seen in Fig. \ref{fig:orbit}. Given that most clones with a semi-major axis
less than 0.5\,pc are stable over 500\,Myr, one quarter of the tidal radius is probably a more reliable metric to measure the long-term stability of Fomalhaut-like systems. 

In Fig. \ref{fig:orbit}, we see that the stable clones in AB+TE simulations could become unstable if Fomalhaut C is included. C tends to perturb B out of rather than into stable orbits because the parameter region corresponding to stable orbits is much smaller than to unstable orbits. Hence the tidal radius for wide binaries cannot be used to measure the stability of triples especially when the separation of the three components are comparable or the decomposition of a triple into two binaries is impossible. From the bottom right panel of Fig. \ref{fig:orbit}, we see that B is more likely to become unstable if C is closer to A because C is more tightly bound to A and thus has a longer time to interact with A and B. 

\begin{figure*}
  \centering
  \includegraphics[scale=0.4]{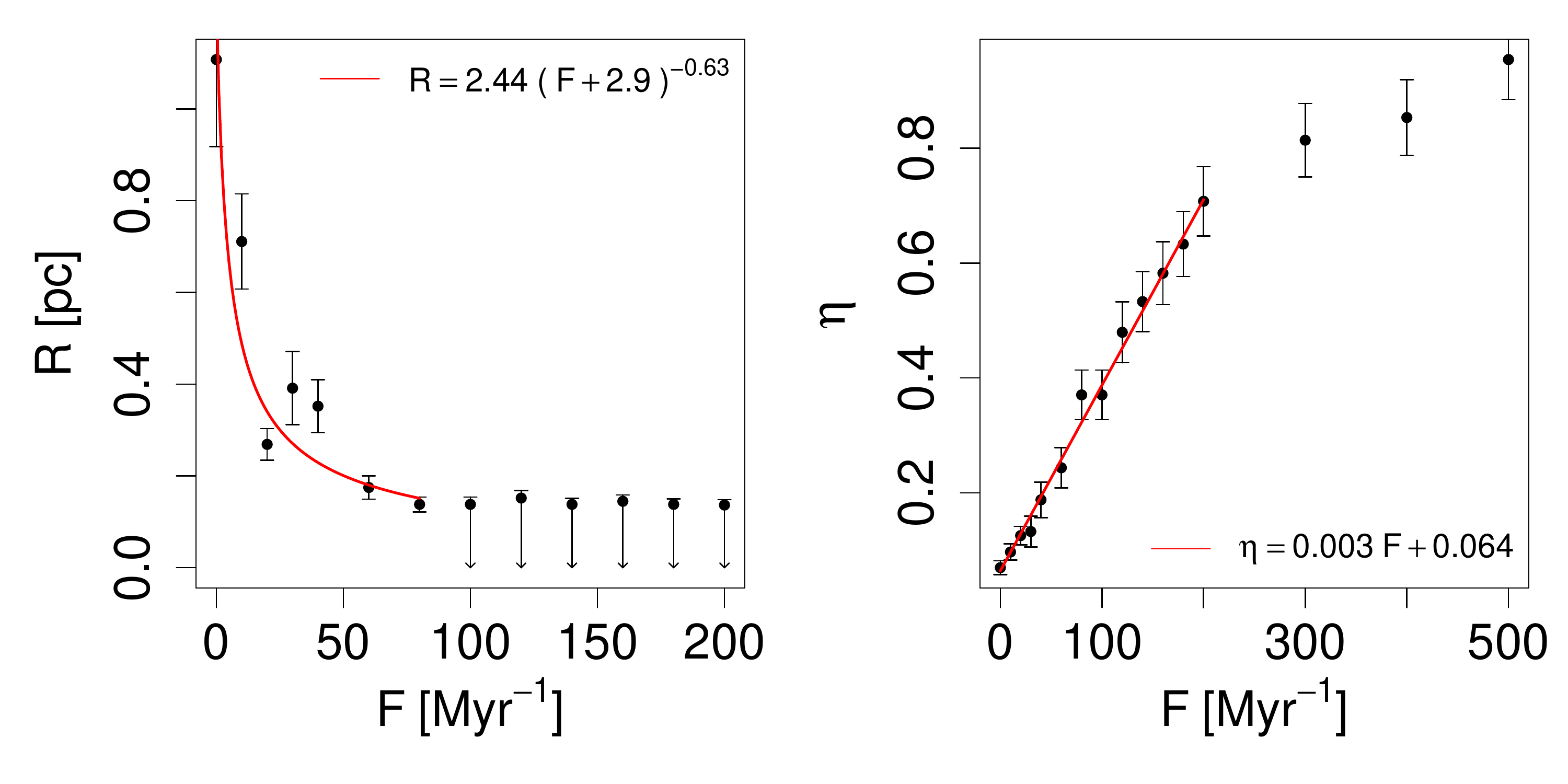}
  \caption{Variation of tidal radius ($R$, left) and ejection ratio ($\eta$, right) with encounter rate $F$. The error bars are calculated assuming a Poisson noise in the number of unstable clones. For saturated simulations, only the upper limits are given. The best-fit functions are shown in the panels. }
  \label{fig:rf}
\end{figure*}

For longer time scales and higher encounter rates, there would be more strong encounters disrupting the system and thus perturbing the system more deeply, leading to a smaller escape radius. To derive an empirical function for escape radius based on the stability analysis of wide binaries, we investigate the dependence of tidal radius on the time scale, encounter rate and other parameters. We define the escape radius as a combined function of integration time $T$ and encounter rate $F$, which is $R(T,F)$. Similarly, we also derive the ejection ratio as a function of $T$ and $F$, which is $\eta(T,F)$. To derive this function, we first fix the integration time at 500\,Myr and simulate 200 clones of Fomalhaut A and B for different encounter rates, ranging from 20 to 500\,Myr$^{-1}$. For each set of simulations, we find the minimum semi-major axis of unstable clones and define it as tidal radius. We show the variation of tidal radius and ejection ratio $\eta$ with encounter rate in Fig. \ref{fig:rf}. We only fit the ``unsaturated'' simulations to deduce the relationship between tidal radius and encounter rate on the basis that the sampling of the parameter space is limited to $>$0.15\,pc and that the tidal radius for the encounter rate $F > 80$ is comparable with the lower limit and thus is ``saturated''. 

We find that the tidal radius is approximately proportional to $F^{-0.63}$. The offset term in the bracket (+2.9) is included to avoid infinite tidal radius for tide-only simulations. Since the encounter rate $F$ is proportional to the cross section $\sigma=\pi q^2$, the minimum periastron $q$ would be proportional to $F^{-0.5}$. However the dependence of tidal radius $R$ on $F$ should be stronger because the tidal radius is determined not only by the strongest encounter but also by cumulative stochastic perturbations from weaker encounters \citep{feng18}. Thus the best-fit shown in the left panel of Fig. \ref{fig:rf} is reasonable and can be applied in the stability analysis of wide binaries separated by more than 0.1\,pc. 

In the right panel of Fig. \ref{fig:rf}, we also see a linear increase of ejection ratio with encounter rate for $F<200$\,Myr$^{-1}$. But for $F>200$\,Myr$^{-1}$, the ejection ratio becomes saturated because few stable clones are left to be disrupted by encounters. Thus we only fit the linear part of the simulations and find $\eta\sim0.003F$. The slope of the fit is also dependent on the masses of the binaries. For example, in the simulations of clones of the Proxima-alpha Centauri system over 7\,Gyr, we find a relationship of $\eta\sim0.005F$. 

\begin{figure}
  \centering
  \includegraphics[scale=0.5]{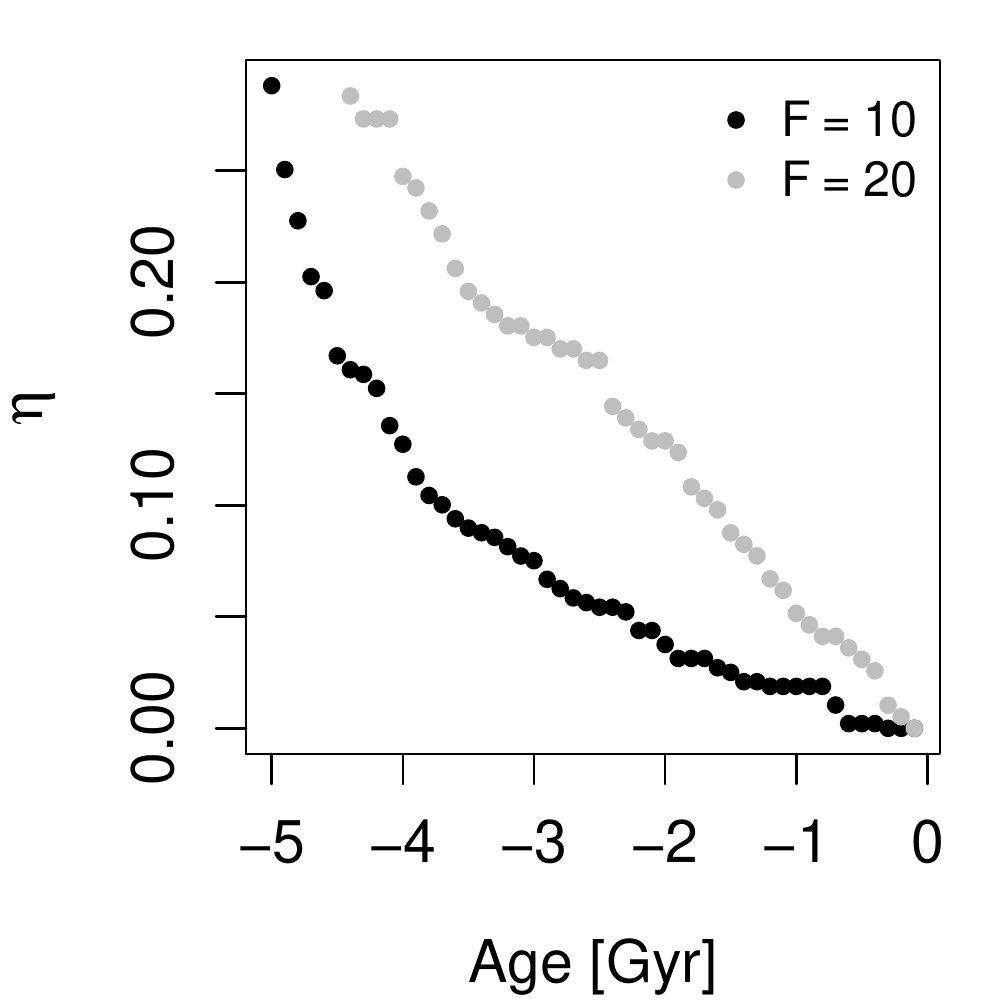}
  \caption{Ejection ratios for simulations with encounter rates of 10 and 20. The simulations are saturated from -5\,Gyr to -3\,Gyr. }
  \label{fig:age}
\end{figure}

Since the perturbations from encounters increase with encounter rate and simulation time span, the encounter rate and the age of a system is degenerate. We test this by showing the ejection ratio for simulations with encounter rates of 10 and 20 over the past 5\,Gyr in Fig. \ref{fig:age}. We see that the slope of the grey dots are approximately double the slope of the black dots, indicating that, a doubling of the time scale is equivalent to a doubling of the encounter rate. Thus the function of escape radius is
\begin{equation}
R~=~2.4~{\rm pc}\left(0.002FT+2.9\right)^{-0.63}~,
\end{equation}
where $T$ is the age of a Fomalhaut-like system. Since the escape radius is proportional to tidal radius, we scale the escape radius to the tidal radius and define the escape radius as
\begin{equation}
  R~=~1.3r_J\left(0.002FT+2.9\right)^{-0.63}~,
\end{equation}
where $r_J\sim 1.7{\rm pc} \left(\frac{M_1+M_2}{2M_\odot}\right)^{1/3}$ is the tidal (Jacobi) radius of a wide binary with masses of $M_1$ and $M_2$ in the solar neighborhood \citep{jiang10}. The tidal force from the Galactic tide becomes significant for binaries separated by more than $r_J$. If stars migrate inward or outward to their current locations \citep{sellwood02,feng13}, the local stellar density and the encounter rate for a star would change with time. This can be accounted for using a more generous escape radius
\begin{equation}
  R~=~1.3r_J\left(0.002\int_0^{T}{F(t)}dt+2.9\right)^{-0.63}~,
\end{equation}
where $F(t)$ is the encounter rate as a function of time.

For example, the tidal radius and escape radius for the Proxima-Alpha Centauri system are respectively 1.71 and 0.03\,pc assuming a constant encounter rate of $F=80$\,Myr$^{-1}$ and an age of 5\,Gyr. The current separation between Proxima and the barycenter of Alpha Centauri is 0.042\,pc \citep{kervella17}, well within the tidal radius but slightly beyond the escape radius. \cite{feng18} show that 17 percent orbits of Proxima's clones become unstable in the 5\,Gyr backward simulations. Therefore the escape radius is more reliable and conservative than the tidal radius in assessing the stability of a binary or multiple system.
%For zero encounter rate, the escape radius is 1.2\,pc, lower than the tidal radius of 1.9\,pc derived from \cite{jiang10}. 

\section{Conclusion}\label{sec:conclusion}
We have assessed the association between Fomalhaut A, B, and C based on the old and new astrometric data and the available radial velocity data. We confirm that the three components are currently associated with each other based on Bayesian model comparison. In pursuit of this result we find a velocity of 8\,km/s for stars separated within 1\,pc as an ``exclusion'' criterion for identification of binary candidates. 

We also study the dynamical evolution of Fomalhaut by simulating the motions of A, B, and C under perturbations from the Galactic tide and stellar encounters. We find that B is stable after the formation of the system with a probability of 66\% by adopting a rate of 80 encounters with periastron less than 1\,pc. Almost all clones of C become unstable during the simulations, apparently supporting the conclusion of \cite{shannon14} that Fomalhaut is currently in a disruption stage. However, a small fraction of C clones are meta-stable over the past 500\,Myr. But since it is rare to observe a triple in disruption, we conclude that Fomalhaut C is in a meta-stable stage rather than in a disruption stage. The former can last for a few hundreds of Myrs while the latter only last for a few Myrs. In other words, Fomalhaut C became bound and unbound repeatedly due to perturbations from the Galactic tide and encounters after the formation of the Fomalhaut system. We propose that Fomalhaut C is a gravitational pair of Fomalhaut A and B. This phenomenon only appears when the orbital energy of wide binaries/multiples is close to zero, making the system extremely sensitive to environment. The gravitational pair provides a new classification when considering the formation and evolution of wide binaries. 

A large sample of Fomalhaut-like systems is required to test the gravitational pair scenario and the Cooper pair analogy. For example, Gaia \citep{gaia16} will provide high precision astrometry for millions of stars which provide six dimensional initial conditions of stars with velocity errors less than 1\,km/s. Numerical integration of the orbits of these stars along with abundance analyses can enable discoveries of coeval and non-coeval gravitational pairs in bound and unbound states.

In various simulations, almost all clones of Fomalhaut A and B with initial semi-major axis larger than half of the tidal radius cannot be stable over 5\,Gyr. Thus the use of tidal radius in stability analysis is misleading probably due to a simplified consideration for encounters. For example, the model of \cite{jiang10} does not account for the peculiar motion of the star, and thus assumes isotropic encounters. This reduces the velocity dispersion of encounters to 40\,km/s, which is less than the value of 50\,km/s based on encounters of the Solar System \citep{rickman08,feng14} and much less than the value of 75\,km/s based on the encounter pairs in the Galaxy \citep{feng17c}. In addition, \cite{jiang10} also assumed a very low stellar number density of 0.05\,pc$^{-3}$ in the solar neighborhood, which is lower than the value derived from the 211 main-sequence stars within 8\,pc of the Sun collected by \cite{kirkpatrick12} and much lower than that in \cite{feng18}. Therefore we suggest one quarter of the tidal radius derived by \cite{jiang10} as a more reliable metric for stability analysis of Fomalhaut-like systems. We also introduce an empirical formula for the escape radius to study the stability of wide binaries with different ages and encounter rates. In the calculation of escape radius, we assume that the Galactic tide and velocity scatter local to the binary is the same as in the solar neighborhood. To avoid such assumptions, a more reliable modeling of stellar encounters and the Galactic tide is essential for stability analysis of wide binaries. 

\section*{Acknowledgement}
This work is supported by the Science and Technology Facilities Council (ST/M001008/1). We used the ESO Science Archive Facility to collect radial velocity data. We thank Eric Mamajek for valuable discussions which help to improve the manuscript. We are very grateful to the anonymous referee for valuable comments.

\bibliographystyle{aasjournal}
\bibliography{nm}
% \end{CJK*}
\end{document}